\documentclass[conference]{IEEEtran}
\IEEEoverridecommandlockouts
\usepackage{cite}
\usepackage{amsmath,amssymb,amsfonts}
\usepackage{graphicx}
\usepackage[labelfont=bf, textfont=bf]{caption}
\usepackage{subcaption}
\usepackage{algorithm}
\usepackage{algpseudocode}
\usepackage{textcomp}
\usepackage{xcolor}
\def\BibTeX{{\rm B\kern-.05em{\sc i\kern-.025em b}\kern-.08em
		T\kern-.1667em\lower.7ex\hbox{E}\kern-.125emX}}

\def\bb0{{\mathbb{0}}}

\def\bb{{\mathbf{b}}}

\def\b0{{\mathbf{0}}}
\def\sf0{{\mathsf{0}}}
\def\rm0{{\mathrm{0}}}
\def\b0{{\pmb{0}}} 
\definecolor{purple(x11)}{rgb}{0.63, 0.36, 0.94}
\definecolor{cadmiumgreen}{rgb}{0.0, 0.42, 0.24}

\begin{document}
	
\title{\texttt{CellSense}: A Sub-6 GHz Cellular ISAC System for Clutter-Robust Passive Sensing
% Object Detection and localization using Sub-6 GHz Cellular signals\\
  \thanks{}
}
	
\author{\IEEEauthorblockN{Bibhor Kumar$^1$, Ish Kumar Jain$^2$, Vijay K Shah$^1$}
  \IEEEauthorblockA{{$^1$\textit{NextG Wireless Lab},} 
    {North Carolina State University},
    Raleigh, North Carolina\\
    {$^2$Department of Electrical, Computer and Systems Engineering,} 
    {Rensselaer Polytechnic Institute},
    Troy, New York\\
    Email: \{bkumar2\}@ncsu.edu, \{jaini\}@rpi.edu, \{vijay.shah\}@ncsu.edu} 
}

\maketitle

\begin{abstract}
Future wireless networks demand capabilities beyond traditional communication, driving the development of Integrated Sensing and Communication (ISAC) for environmental awareness, localization, and tracking. Ubiquitous cellular deployment allows ISAC to maximize spectral efficiency, lower costs, and expand sensing coverage. However, sub-6 GHz research has heavily favored communication, leaving sensing capabilities largely under-explored. To bridge this gap, we introduce \texttt{CellSense}, a novel sub-6 GHz ISAC architecture natively integrated into the 5G cellular protocol stack for real-world target tracking. We validate the system via Sionna-based orthogonal frequency-division multiplexing (OFDM) link-level simulations and an experimental USRP hardware prototype using the OpenAirInterface (OAI) stack. Furthermore, we analyze the communication–sensing tradeoff by quantifying how pilot symbol density impacts throughput versus sensing accuracy. Simulations show that \texttt{CellSense} achieves a $74\%$ detection probability with a $1.43$ m localization error in indoor warehouse environment, which improves to $94\%$ detection and a sub-meter error of $0.33$ m in the outdoor environment of Oval area at the NCSU Centennial campus. Hardware experiments in a highly cluttered indoor laboratory confirm a $1.28$ m localization accuracy and $76\%$ detection probability, proving its efficacy for practical ISAC deployments.
% \textcolor{red}{[talk all three 3GPP ISAC metrics]}
% \textcolor{red}{hardware prototype-based} 
% \textcolor{red}{[also, talk about outdoor simulation-based centennial campus]}
\end{abstract}

\section{Introduction}
ISAC has emerged as a cornerstone of next-generation wireless systems, recognized by both the ITU \cite{itur_m2160_2023} and 3GPP \cite{3gpp_tr_22837_2023} as one of the six primary usage scenarios for 6G. This paradigm is driven by its ability to dual-purpose wireless networks for both sensing and communication without requiring additional spectrum or dedicated hardware resources. Due to these capabilities, ISAC is poised to deliver critical utility across both commercial and defense domains, enhancing automated driving and smart cities on one hand, while maximizing tactical situational awareness on the other. To standardize these capabilities, the 3GPP Release 19 specifications \cite{3gpp_ts_22137_2024} outline several distinct sensing service categories. Among these, \textit{object detection and tracking} stands out as a particularly vital and practical use case, and it is this specific application of ISAC that forms the central focus of our work. \par

Despite its potential, repurposing current cellular systems for ISAC introduces two major limitations. First, communication-centric OFDM waveforms are unoptimized for high-precision parameter estimation, which restricts sensing resolution and accuracy. Second, while standard OFDM pilots are designed solely for communication CSI acquisition at the BS and UE, high-performance sensing demands much denser pilot transmissions. This introduces a fundamental communication–sensing tradeoff, where maximizing sensing accuracy directly reduces communication throughput. \par

To address these limitations, we present \texttt{CellSense}, a high-precision ISAC system based on OFDM signals designed for real-time target localization and tracking within existing Sub-6 GHz cellular infrastructure. Implemented on a full-stack 5G NR testbed, \texttt{CellSense} demonstrates robust cellular sensing by utilizing standard-compliant pilot signals without requiring modifications to the underlying communication protocol. In summary, our primary contributions are:
 \begin{itemize}
 \item We propose \texttt{CellSense}, a comprehensive Sub-6 GHz ISAC architecture that is fully integrated with a functional 5G cellular stack.
 
 \item We design an iterative parameter estimation and refinement method that isolate individual multipath components for improved estimation accuracy. This is paired with a robust data association and tracking pipeline that utilizes multi-frame sliding windows and bipartite matching to maintain persistent target tracks in the presence of heavy environmental clutter. 
    
    % \textcolor{red}{[Rewrite this sentence to make it more clear how CellSense architecture was designed to enable high-precision sensing accuracy.]}
 \item Finally, we provide extensive simulation and hardware prototype based validation of \texttt{CellSense} across multiple environments, demonstrating high-fidelity sensing with minimal communication overhead. 
 %Compared to the baseline, CellSense improves detection probability ($P_d$) by $52\%$ and reduces localization RMSE by $58.3\%$ for indoor simulation environment; outdoor simulations show a $10\%$ $P_d$ increase and an $86.1\%$ error reduction to a sub-meter $0.33$ m accuracy. Our hardware experiments demonstrate a $33\%$ improvement in $P_d$ alongside a $41.8\%$ reduction in localization RMSE, confirming its real-world efficacy. Furthermore, quantifying the communication–sensing tradeoff reveals that a dense pilot configuration optimizes tracking accuracy to $0.66$ m with minimal throughput penalty.

 \end{itemize}
    
    % \textcolor{red}{[Give exact numbers for all three 3GPP sensing metrics, and sensing-comm tradeoff for your evaluation scenarios.]}

% The remainder of this paper is organized as follows, Sec. II discusses related work. Sec. III describes our system and signal model. Sec. IV describes \texttt{CellSense} architecture in extensive detail. Sec. V provide simulation and hardware experiments based performance evaluation. Sec. VI concludes the paper. 

%\textit{Paper Organization.} Section II presents the system and signal model. Section III details the proposed CellSense system. Sections IV provides the simulation and hardware experimental results, and Section V concludes the paper.

% What is the problem? (clear problem definition)\\

% Why is it hard? (interesting, innovative, novel, etc)\\

% What is our intuition? (key insights, observation, addressing existing gaps)\\

% What is our solution? (challenges and solutions)\\

% \begin{itemize}
% \item First paragraph will be for motivation of a cellular ISAC system
% \item In second and third paragraphs, we will provide a summary of previous related works on ISAC in cellular networks, WiFi networks including both simulation based papers and hardware prototypes based papers.
% \item Last 1-2 paragraphs will describe our contributions and section-wise breakdown of the paper.
% \end{itemize}

\section{Related Work}
The dual challenges of waveform optimization and the communication–sensing trade-off have catalyzed extensive research in ISAC. Several foundational studies \cite{wei20225g, demir2024scout, sagduyu2025multi, ma2022downlink} have demonstrated the theoretical feasibility of repurposing standard 5G reference signals such as the Positioning Reference Signal (PRS) or the Synchronization Signal Block (SSB) for sensing applications. However, such studies predominantly rely on idealized channel simulations that neglect non-linear hardware impairments, phase noise, and intricate multipath scattering.
To bridge this gap, Software-Defined Radio (SDR) experimental testbeds have emerged to evaluate standard-compliant 5G New Radio (NR) signals under realistic operational conditions \cite{li2025msac, yang2024isac, pegoraro2024hisac, wild20236g}. However, existing hardware prototypes suffer from two primary limitations: they prioritize mmWave frequencies over the ubiquitously deployed Sub-6 GHz bands, and they operate in isolation rather than integrating with a functional, full-stack 5G cellular network. In contrast, \texttt{CellSense} is explicitly designed for Sub-6 GHz frequencies and seamlessly integrates with a functional 5G cellular stack, offering the practical deployability that current testbeds lack.
Parallel to cellular research, Wi-Fi-based ISAC systems \cite{meng2023secur, he2025versabeam, xie2019md, he2023sencom, kotaru2015spotfi, joshi2015wideo} have been proposed as low-cost alternatives for localized applications, leveraging existing IEEE 802.11 signals. However, their low transmit power restrictions and structural indoor limitations inherently restrict their suitability for high-mobility, long-range outdoor sensing scenarios. \par

\begin{figure}[t!]
\centering
\includegraphics[width=0.9\linewidth]{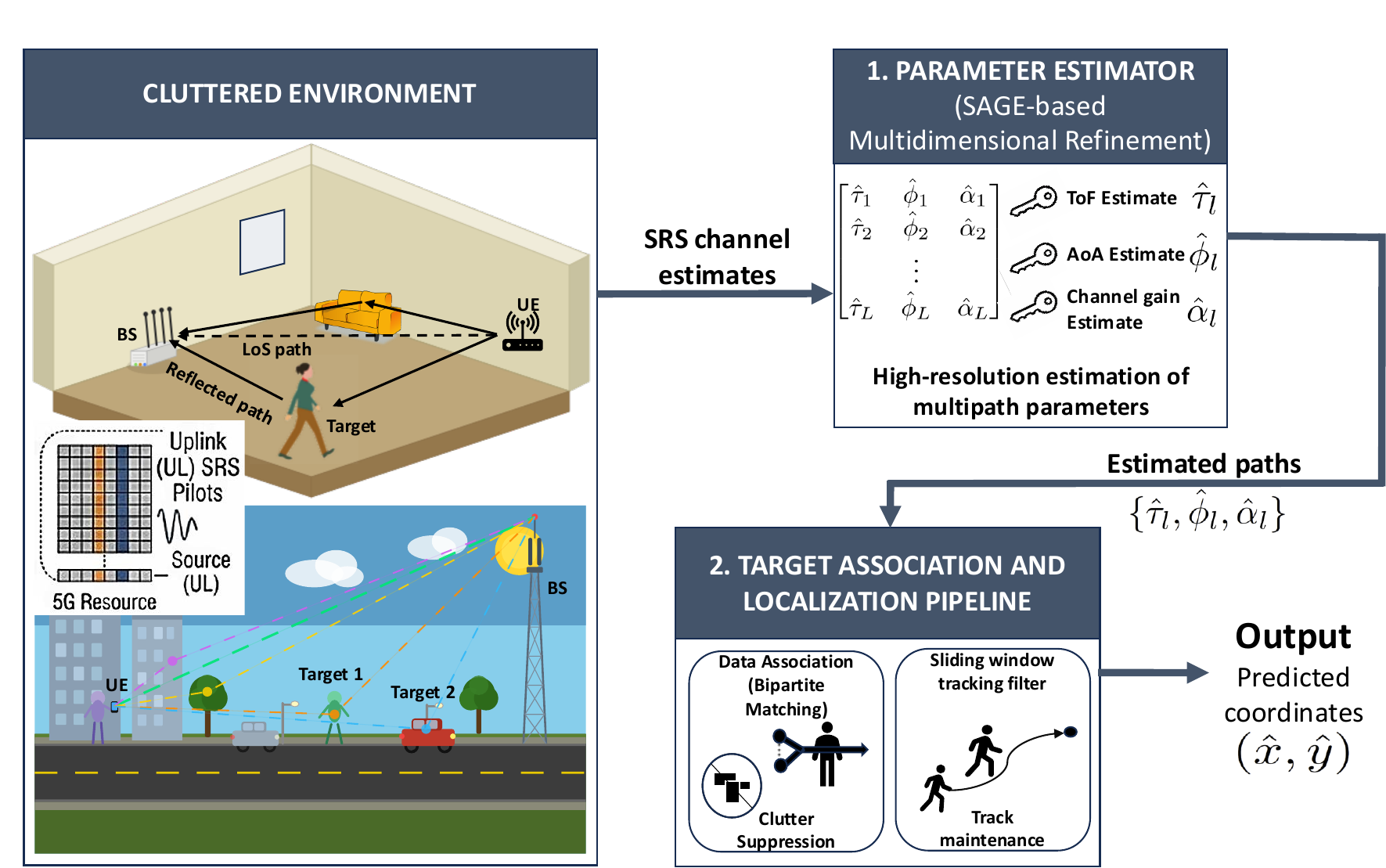}
\vspace{-1.5mm}
\caption{Overview of the \texttt{CellSense} architecture}
\label{fig:CellSense}
\vspace{-7mm}
\end{figure}

\section{System model and signal model}
\texttt{CellSense} consists of a Single-Input Multiple-Output (SIMO) system, featuring a multi-antenna 5G BS equipped with a Uniform Linear Array (ULA) and a single-antenna UE. The system, depicted in Fig. \ref{fig:CellSense}, operates in a cluttered environment where a dynamic target must be detected and tracked amidst environmental reflections. Both the BS and UE are running full 5G protocol stack and maintain a continuous communication link between them. We further assume that the BS and UE are stationary and their locations are known a priori. This assumption is reasonable in multi-BS deployments, where UE localization can be obtained using GPS or through measurements from multiple BSs. In addition, we assume that a line-of-sight (LoS) link between the UE and BS is always available. For channel estimation, we exploit uplink Sounding Reference Signal (SRS) pilot transmissions from the UE for uplink channel state information (CSI) estimation. The SRS pilot signals are based on Zadoff-Chu sequences \cite{3gpp_ts_38211_2020}. We adopt an uplink-based sensing framework to leverage the higher computational capability at the BS, enabling centralized processing for object detection and tracking tasks. \par
The received signal at the BS can be modeled as
\vspace{-2mm}
\begin{align}
    \mathbf{Y}_m(t) = \sum_{l=0}^{L-1} \alpha_l e^{j2 \pi \gamma_l t} \mathbf{a}_{BS}(\phi_l) \mathbf{x}_m^T(t - \tau_l) &p(t - \tau_l - mT) \nonumber \\ 
    \vspace{-5mm}
    &+ \mathbf{N}(t),
    \label{eq:signal_model}
\end{align}
where $\alpha_l, \gamma_l, \tau_l$ and $\phi_l$ represent the complex pathgain, Doppler, transmit delay and angle of arrival (AoA) of the $l^{th}$ received path at the BS with $L$ being the total number of paths. Also, $\mathbf{a}_{BS}(\phi_l) = [1, \ e^{j2 \pi d sin(\phi_l)}, ..., \ e^{j2 \pi (M-1) d sin(\phi_l)}]$ represents the antenna array response with $M$ being the no. of antennas and $d$ being the relative inter-antenna distance with respect to the wavelength at the BS. Also, $p(t)$ represents the pulse shaping function, $\mathbf{N}(t)$ represents the receiver noise, $T$ is the OFDM symbol duration and $m$ is the OFDM symbol number. Finally, $\mathbf{x}_m(t)$ represents the transmitted OFDM signal modeled as $\mathbf{x}_m(t) = \Sigma_{k=0}^{N_{fft}-1} X_{k,m} e^{j2 \pi k f_{scs}(t-n_{cp}/f_s)}.$ Here, $X_{k,m}$ is the symbol transmitted on $k^{th}$ subcarrier of $m^{th}$ OFDM symbol, $f_{scs}$ is the subcarrier spacing, $f_s$ is the sampling rate and $n_{cp}$ is the length of the cyclic prefix (CP). 

\textbf{Channel estimation:} Upon receiving the signal, the BS OFDM receiver samples the continuous-time waveform. The resulting time-domain samples, $\mathbf{Y}_m(n/f_s)$ for $n \in \{n_{cp}, n_{cp} + 1, \dots, N_{fft} + n_{cp}-1\}$, are transformed into the frequency domain via Discrete Fourier Transform (DFT), yielding the frequency-domain symbols $\mathbf{Y}_m^f[k]$ for $k \in \{0, 1, \dots, N_{fft}-1\}$. After that, we perform least squares (LS) estimation using the pilot symbols to estimate the channel for the corresponding subcarriers given as \cite{malik2024concept}
\vspace{-2mm}
\begin{align}
    \hat{\mathbf{H}}_{m}^f[q][k] = \frac{\mathbf{Y}_m^f[q][k]}{X^p_{k,m}}.
\end{align}
Here, $k$ denotes the pilot subcarrier, $m$ denotes the OFDM symbol, $q$ denotes the receive antenna index, $X^p_{k,m}$ denotes the corresponding pilot symbol. $\mathbf{Y}_m^f[q][k]$ denotes the received symbol and $\hat{\mathbf{H}}_{m}^f[q][k]$ denotes the channel estimate for $k^{th}$ subcarrier and $q^{th}$ receive antenna. To maximize throughput efficiency, SRS pilots are mapped using a comb structure with a comb size of two or greater. While this configuration enables the multiplexing of data and SRS symbols within the same OFDM resource element grid, it inherently yields sparse channel estimates. Linear interpolation is subsequently applied across the frequency domain to estimate the channel coefficients for the remaining unpiloted subcarriers \cite{malik2024concept}.

\section{CellSense Design}
In this section, we will discuss the individual components of the \texttt{CellSense} architecture in more detail. As illustrated by the functional flow diagram in Fig. \ref{fig:CellSense}, the proposed system is designed to seamlessly integrate high-precision passive sensing into standard 5G communication frameworks. The architecture can be functionally decomposed into two primary, sequentially linked stages: \textit{Parameter Estimation}, and \textit{Target Association and Localization}. Each of these stages is described in detail in the following subsections.

\subsection{Stage 1: Parameter estimation} % Provide the details of the algorithm for parameter estimation here. Do not mention that you algorithm is based on mD-Track, say that it is based on SAGE algorithm.
After getting the SRS channel estimates from the OFDM receiver, \texttt{CellSense} extracts the fundamental physical parameters of the propagation environment, specifically the Angle of Arrival (AoA) and the propagation delay. We begin by reconstructing the received signal using the provided channel estimates and a pilot sequence as $\hat{\mathbf{Y}}_m^f[q][k] = \hat{\mathbf{H}}_m^{f}[q][k] X^p_{k}.$ Here $\hat{\mathbf{Y}}_m^f[q][k]$ is the generated received signal for $k^{th}$ subcarrier and $q^{th}$ antenna and $X^p_k$ is the pilot symbol for $k^{th}$ subcarrier. This frequency-domain representation is subsequently transformed into the time domain via an Inverse Discrete Fourier Transform (IDFT), yielding $\hat{\mathbf{Y}}_m = IDFT(\hat{\mathbf{Y}}_m^f).$ \par
Inspired by the Wi-Fi-based mD-Track algorithm \cite{xie2019md}, we resolve the multipath components from the reconstructed time-domain signal by employing an iterative algorithm based on the Space-Alternating Generalized Expectation-maximization (SAGE) framework \cite{fessler2002space}. The SAGE algorithm effectively decomposes the multi-dimensional maximum likelihood estimation problem into a sequence of one-dimensional optimization tasks. The expectation step (E-step) is characterized by the correlation function:
\vspace{-1mm}
\begin{align}
    z(\phi, \tau) = \sum_{n=0}^{N_{fft}-1} \mathbf{a}_{BS}^H(\phi) \hat{\mathbf{Y}}_m[:, n] \mathbf{X}_p^{*}[n-\tau],
    \label{eq:expectation}
\end{align}
where $\mathbf{X}_p^{*}$ denotes the complex conjugate of the time-domain pilot sequence used in the signal regeneration. The subsequent maximization step (M-step) identifies the parameter pair that maximizes this correlation:
\vspace{-2.5mm}
\begin{align}
    (\hat{\phi}, \hat{\tau}) = \underset{(\phi, \tau)}{\text{argmax}} \  |z(\phi, \tau)|.
    \label{eq:maximization}
\end{align}
Combining Eq. \ref{eq:signal_model} and Eq. \ref{eq:expectation} with the maximization step, we can estimate the pathgain as $\hat{\alpha} = z(\hat{\phi}, \hat{\tau})/(M \times N_{fft}).$ \\
While these operations are rooted in the Generalized EM (GEM) architecture, our approach diverges by estimating multipath parameters in a successive, path-by-path manner. Upon completing the initial E-step and M-step to resolve the parameters of the dominant propagation path, the contribution of this path is reconstructed and subtracted from the total signal $\hat{\mathbf{Y}}_m$ to yield a residual signal:
\vspace{-2mm}
\begin{align}
    \hat{\mathbf{W}} = \hat{\mathbf{Y}}_m - \hat{\alpha} \mathbf{a}_{BS}(\hat{\phi}) \mathbf{X}_p^{T}[n-\hat{\tau}].
    \label{eq:residual}
\end{align}
This iterative signal cancellation process is repeated on the residual signal $\hat{\mathbf{W}}$ to extract secondary and weaker multipath components. This cycle continues until a predefined number of paths have been estimated and final residual signal is calculated as
\vspace{-4mm}
\begin{align}
    \hat{\mathbf{W}} = \hat{\mathbf{Y}}_m - \sum_{l=0}^{L_{est}-1} \hat{\alpha}_l \mathbf{a}_{BS}(\hat{\phi}_l) \mathbf{X}_p^{T}[n-\hat{\tau}_l].
\end{align}

The initial stage of the algorithm provides a set of coarse estimates for the $L_{est}$ dominant propagation paths. However, in dense multipath environments typical of tactical scenarios, the accuracy of these initial estimates can be degraded by mutual interference between paths-particularly when components are closely spaced in the delay or angular domains. To mitigate this "leakage" effect, our proposed algorithm performs a refinement phase to achieve higher estimation precision. 
The refinement process begins by isolating each path from the aggregate signal. For the $l$-th path, we reconstruct a ``cleaned" version of the received signal, $\hat{\mathbf{Y}}_l$, by combining the final residual signal $\hat{\mathbf{W}}$ with the previously estimated contribution of that specific path:
\vspace{-1.5mm}
\begin{align}
    \hat{\mathbf{Y}}_l = \hat{\mathbf{W}} + \hat{\alpha}_l \mathbf{a}_{BS}(\hat{\phi}_l) \mathbf{X}_p^{T}[n-\hat{\tau}_l],
\end{align}
where $\hat{\phi}_l$, $\hat{\alpha}_l$ and $\hat{\tau}_l$ represent the coarse estimates for the path of interest. We then re-apply the SAGE-based maximization steps (Eq. \ref{eq:expectation} and Eq. \ref{eq:maximization}) to this reconstructed signal $\hat{\mathbf{Y}}_l$ to derive updated, high-precision parameters. Because the interference from all other $L_{est}-1$ paths has been suppressed in the residual $\hat{\mathbf{W}}$, this iteration significantly reduces the bias introduced during the initial coarse estimation. Following the update of the $l$-th path's parameters, the residual signal $\hat{\mathbf{W}}$ is recalculated using Eq. \ref{eq:residual}. This refinement cycle is performed sequentially for all estimated paths and repeated for a predetermined number of iterations, ensuring that the final parameters $(\hat{\phi}_l, \hat{\tau}_l, \hat{\alpha}_l)$ converge toward the true maximum likelihood estimates.

\begin{figure}%
    \centering
    \subfloat[\centering Detecting the target]{{\includegraphics[width=3.5cm]{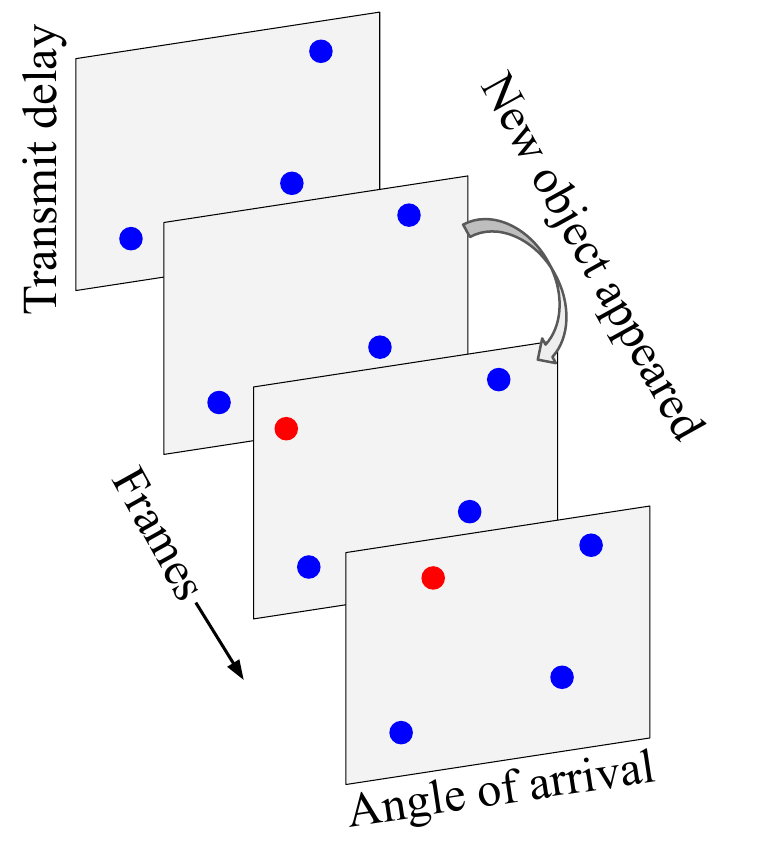} }}%
    \qquad
    \subfloat[\centering Tracking the target]{{\includegraphics[width=3.5cm]{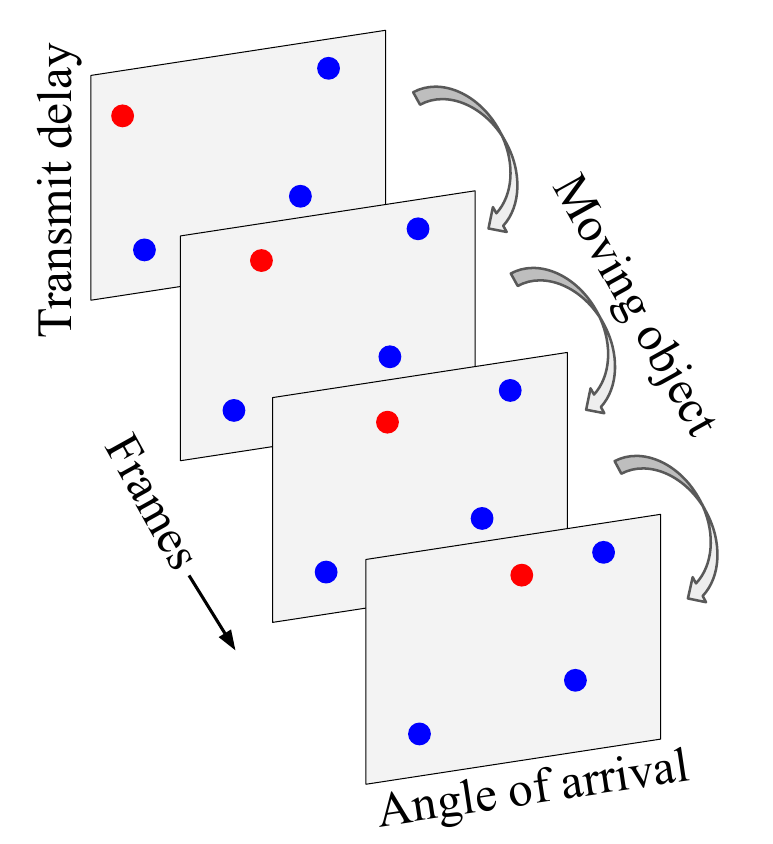} }}%
    \vspace{-1.5mm}
    \caption{Heuristic methods for dynamic object detection and tracking using \texttt{CellSense}. Blue and red markers indicate static reflectors and the dynamic target, respectively.}%
    \label{fig:object_detect}%
    \vspace{-5mm}
\end{figure}

% \begin{table*}[t]
% \centering
% \caption{Simulation results: Detection and localization performance comparison for CellSense and 2D-FFT}
% \label{table:simulation_results}
% \begin{tabular}{|l|l|l|l|l|l|l|l|l|l|}
% \hline
%      & \multicolumn{3}{|c|}{\textbf{Detection probability}} & \multicolumn{3}{|c|}{\textbf{False alarm probability}} & \multicolumn{3}{|c|}{\textbf{Localization RMSE (meters)}} \\ \hline
% & Indoor & Outdoor & Open & Indoor & Outdoor & Open & Indoor & Outdoor & Open \\ \hline

% \textbf{CellSense} & 0.74 & 0.94 & 1 & 0.10 & 0.12 & 0.02 & 1.43 & 0.33 & 0.28 \\ \hline

% \textbf{2D-FFT} & 0.22 & 0.84 & 0.88 & 0.88 & 0.42 & 0.00 & 3.43 & 2.38 & 1.65 \\ \hline

% \end{tabular}
% \end{table*}

\subsection{Stage 2: Target association and localization} 
Following Stage 1 (parameter estimation), the resolved multipath components need to be associated with distinct physical entities, enabling the separation of dynamic targets from static environmental clutter. To accomplish this, in Stage 2, we employ two complementary methodologies for target detection and tracking, as illustrated in Fig. \ref{fig:object_detect}. \par
The first approach employs \textit{Differential Change Detection} which assumes that the initial sensing environment is static and can therefore serve as a baseline channel profile. Any significant parameter deviation from this baseline indicates a dynamic object entering the sensing region, enabling the system to autonomously detect new targets by continuously monitoring variations relative to the baseline. The second approach relies on \textit{Temporal Trajectory Analysis} across consecutive sensing frames. By examining parameter evolution across consecutive frames, the system distinguishes and suppresses persistent static clutter from moving reflections, thereby maintaining continuous tracks of mobile targets over time. Together, these complementary heuristics provide robust target discrimination and reliable tracking performance, even in densely cluttered environments. \par
\textbf{Parameter-reflector association:} A fundamental challenge in persistent tracking is the consistent association of resolved parameters (AoA and transmit delay) with their respective physical reflectors across consecutive temporal snapshots. However, in dense scattering environments, weaker reflections may fluctuate above and below the detection threshold, leading to intermittent parameter estimates across frames. To ensure temporal continuity, \texttt{CellSense} maintains a global repository of parameters spanning a sliding window of ten consecutive snapshots. We formulate parameter-to-reflector mapping as a combinatorial assignment problem, minimizing the aggregate distance between current parameters and established tracks in the global repository. By computing pairwise distances between all points of the current snapshot and global repository, we solve this bipartite matching problem to identify the optimal assignment. \par

\begin{figure}[t!]
    \centering
    \begin{subfigure}[b]{0.48\columnwidth} 
        \centering
        \includegraphics[width=\linewidth]{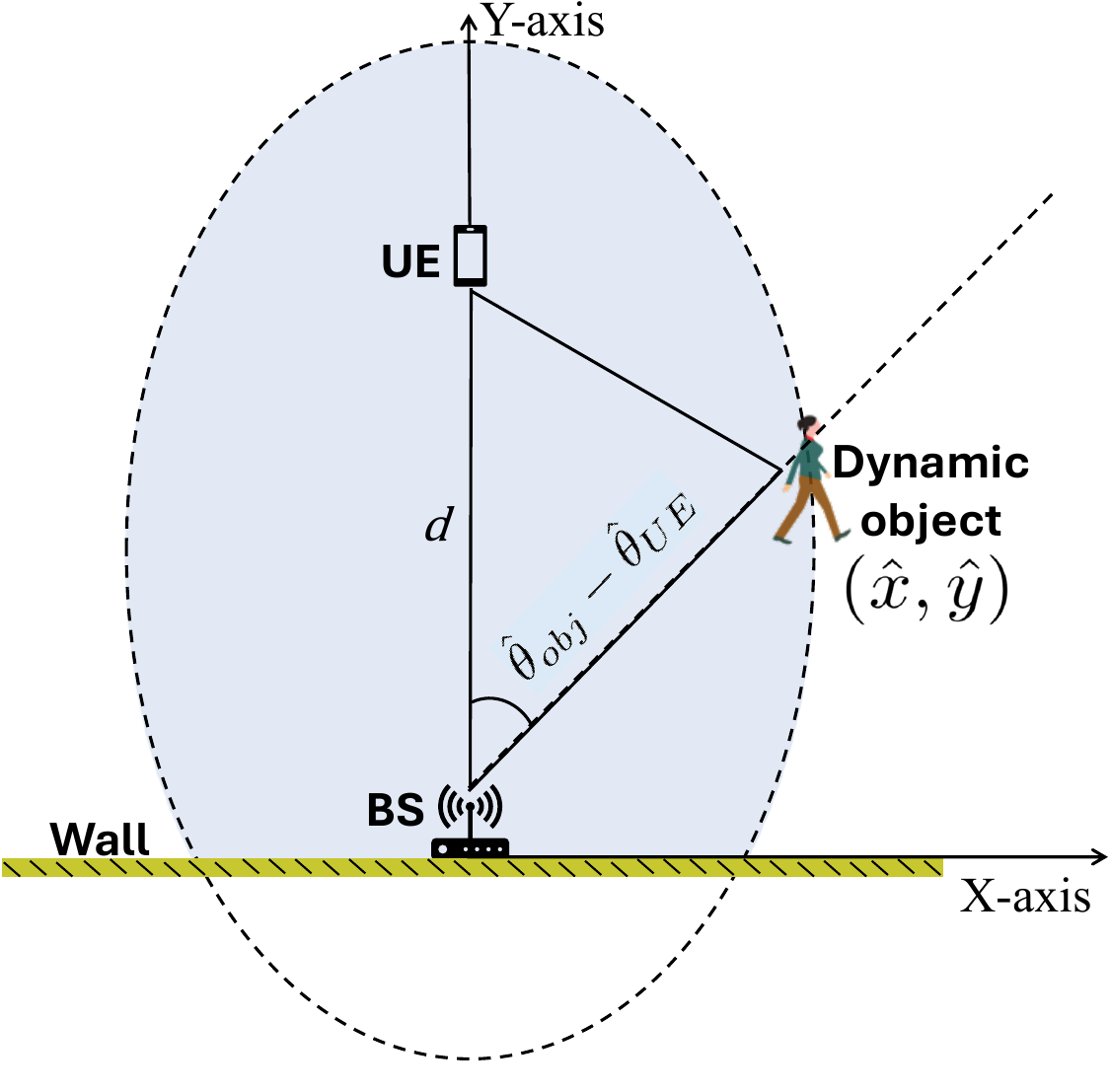}
        \caption{}
        \label{fig:Localization}
    \end{subfigure}
    \begin{subfigure}[b]{0.42\columnwidth}
        \centering
        \includegraphics[width=\linewidth]{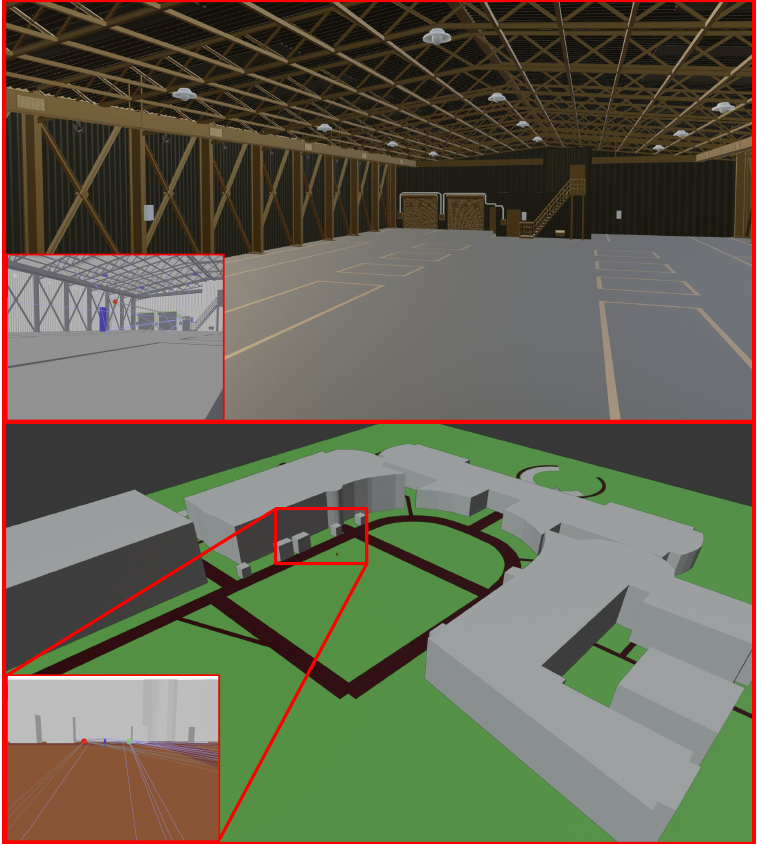}
        \caption{}
        \label{fig:sim_env}
    \end{subfigure}
    \vspace{-2mm}
    \caption{(a) Localization of a dynamic target via the intersection of a bistatic range ellipse (with BS and UE as foci) and a directional ray originating from the BS. (b) Site-specific Sionna ray-tracing environments generated in Blender: indoor warehouse (top) and NCSU Centennial Campus outdoor Oval (bottom).}
    \label{fig:sensing_metrics}
    \vspace{-6mm}
\end{figure}

\textbf{Localization:} Once the dynamic parameters are isolated from the environmental multipath, the final step is to localize the object in a 2-D coordinate system. This process leverages the known location of the UE and the assumption of a persistent Line-of-Sight (LoS) connection between the BS and UE. The LoS component is readily identified as the reference signal due to its known angular direction and dominant power profile. By utilizing the differential delay between the target-reflected path and the direct LoS path, we estimate the sum of the distances from the object to both the BS and the UE as $\hat{l} = c \times (\hat{\tau}_{obj} - \hat{\tau}_{UE}) + d$. Here $c$ is the speed of radio waves in air, $\hat{\tau}_{obj}$ and $\hat{\tau}_{UE}$ are estimated transmit delays from the object and UE, respectively and $d$ is the known BS-UE distance. The moving object should be on the intersection of the ellipse with UE and BS as foci and sum of distance from the 2 foci as $\hat{l}$ and a ray originating from the BS at an angle of $\hat{\theta}_{obj} - \hat{\theta}_{UE}$ relative to the BS-UE LoS baseline, as shown in Fig. \ref{fig:Localization}. Here, $\hat{\theta}_{obj}$ and $\hat{\theta}_{UE}$ are estimated AoA from object and UE, respectively. Using these information, the 2D location estimate of the object of interest is given as
\vspace{-2mm}
\begin{align}
    &\hat{x} = \frac{\left[  \frac{\hat{l}^2 sin^2(\hat{\theta}_{obj} - \hat{\theta}_{UE})}{2 (\hat{l}cos(\hat{\theta}_{obj} - \hat{\theta}_{UE}) - d)} + \frac{\hat{l}cos(\hat{\theta}_{obj} - \hat{\theta}_{UE}) + d}{2} \right]} {\left[ \frac{\hat{l} sin(\hat{\theta}_{obj} - \hat{\theta}_{UE})}{\hat{l}cos(\hat{\theta}_{obj} - \hat{\theta}_{UE}) - d} + cot(\hat{\theta}_{obj} - \hat{\theta}_{UE}) \right]} \nonumber \\ 
    & \ \ \ \ \ \ \ \ \ \ \ \ \ \hat{y} = cot(\hat{\theta}_{obj} - \hat{\theta}_{UE}) \times \hat{x}.
    \label{eq:localization}
\end{align}

\begin{figure}[t!]
\centering
\includegraphics[width=0.8\linewidth]{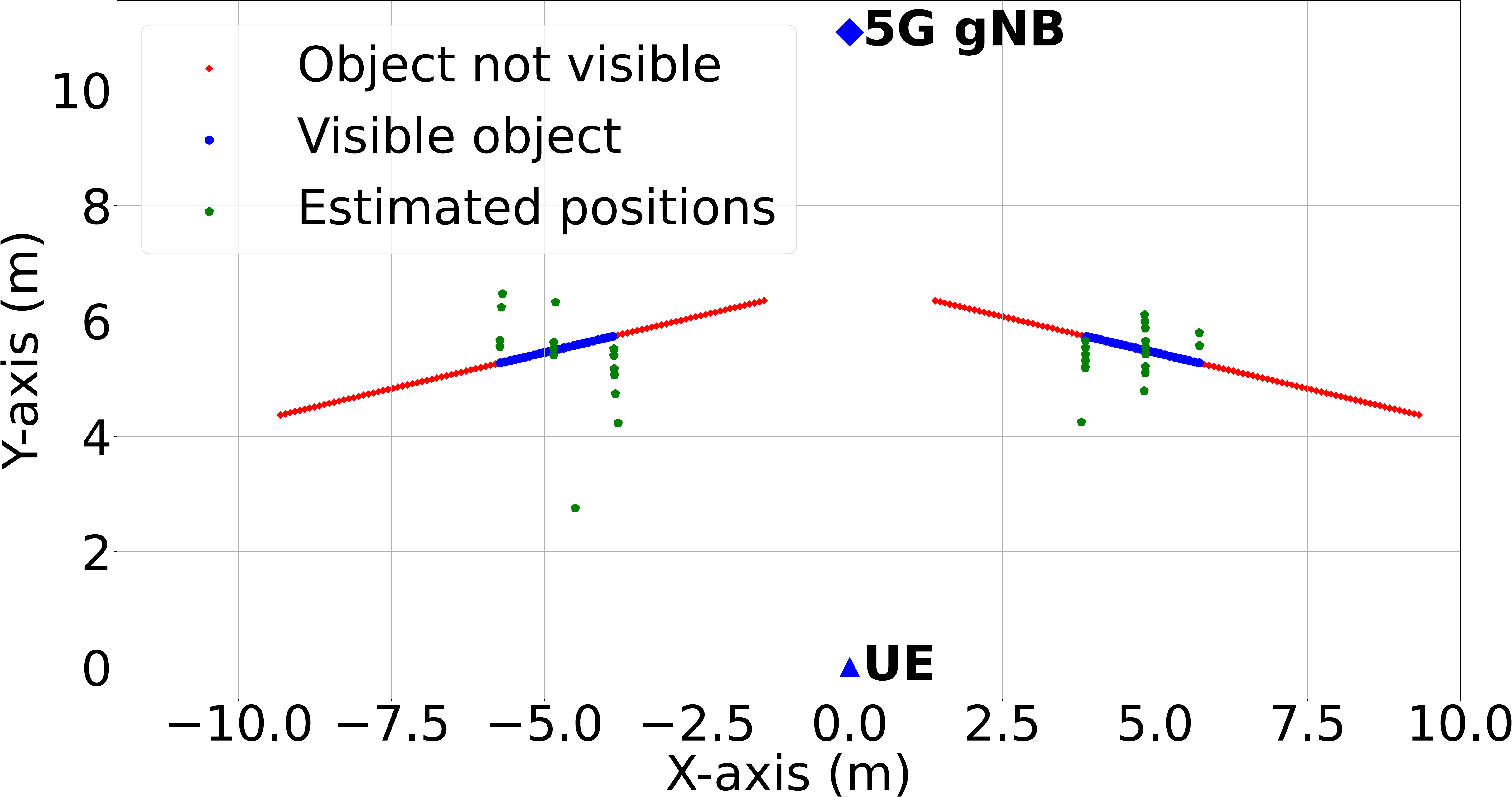}
\vspace{-0.05in}
\caption{Comparison between two ground truth trajectories and the estimated coordinates for the target of interest in the outdoor simulation environment.}
\label{fig:trajectories}
\vspace{-6mm}
\end{figure}

\begin{figure*}[t!]
    \centering
    \begin{subfigure}[b]{0.3\textwidth}
        \centering
        \includegraphics[width=\linewidth]{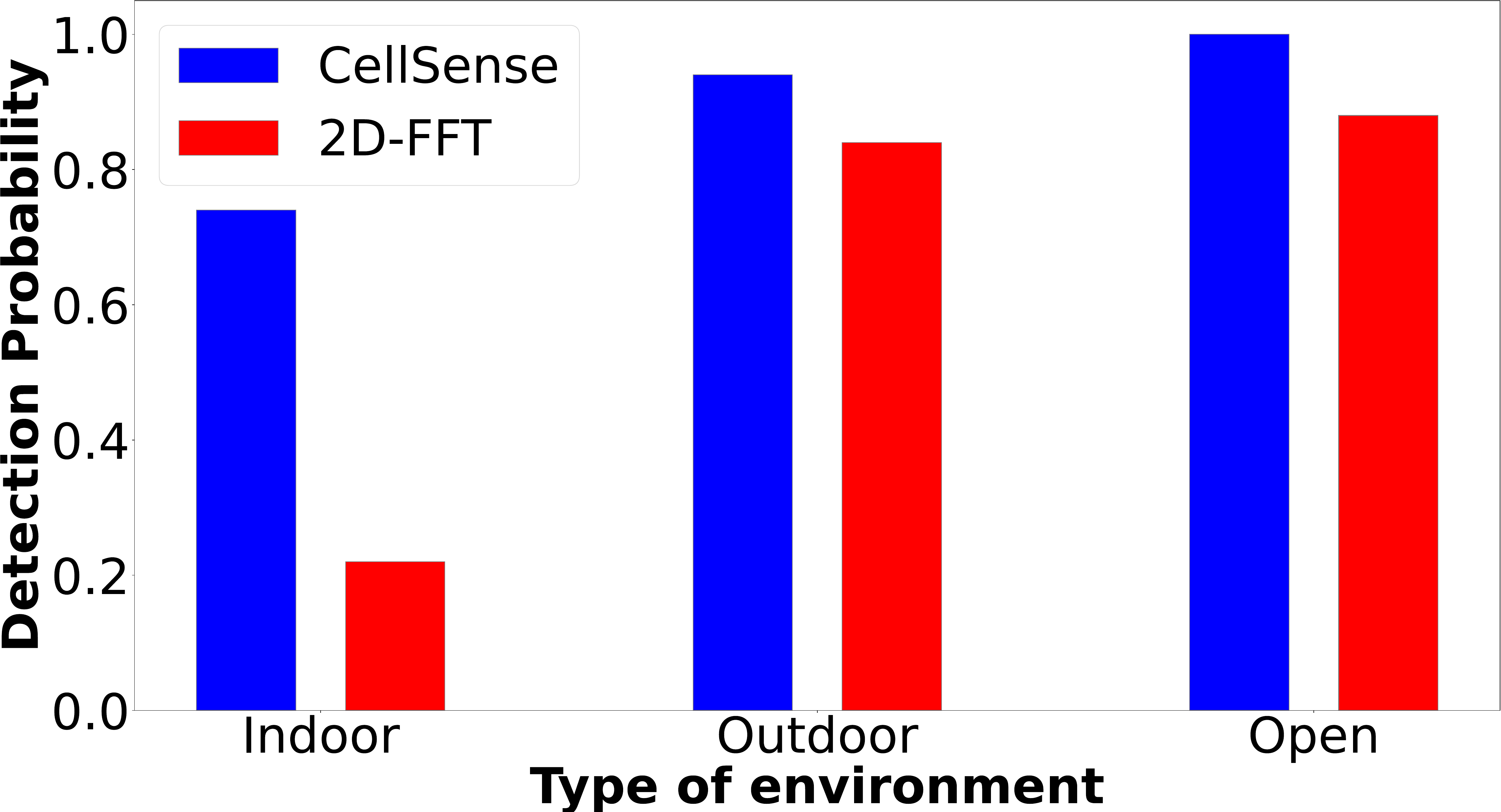}
        \caption{Detection probability}
        \label{fig:sub_image1}
    \end{subfigure}
    \hfill
    \begin{subfigure}[b]{0.3\textwidth}
        \centering
        \includegraphics[width=\linewidth]{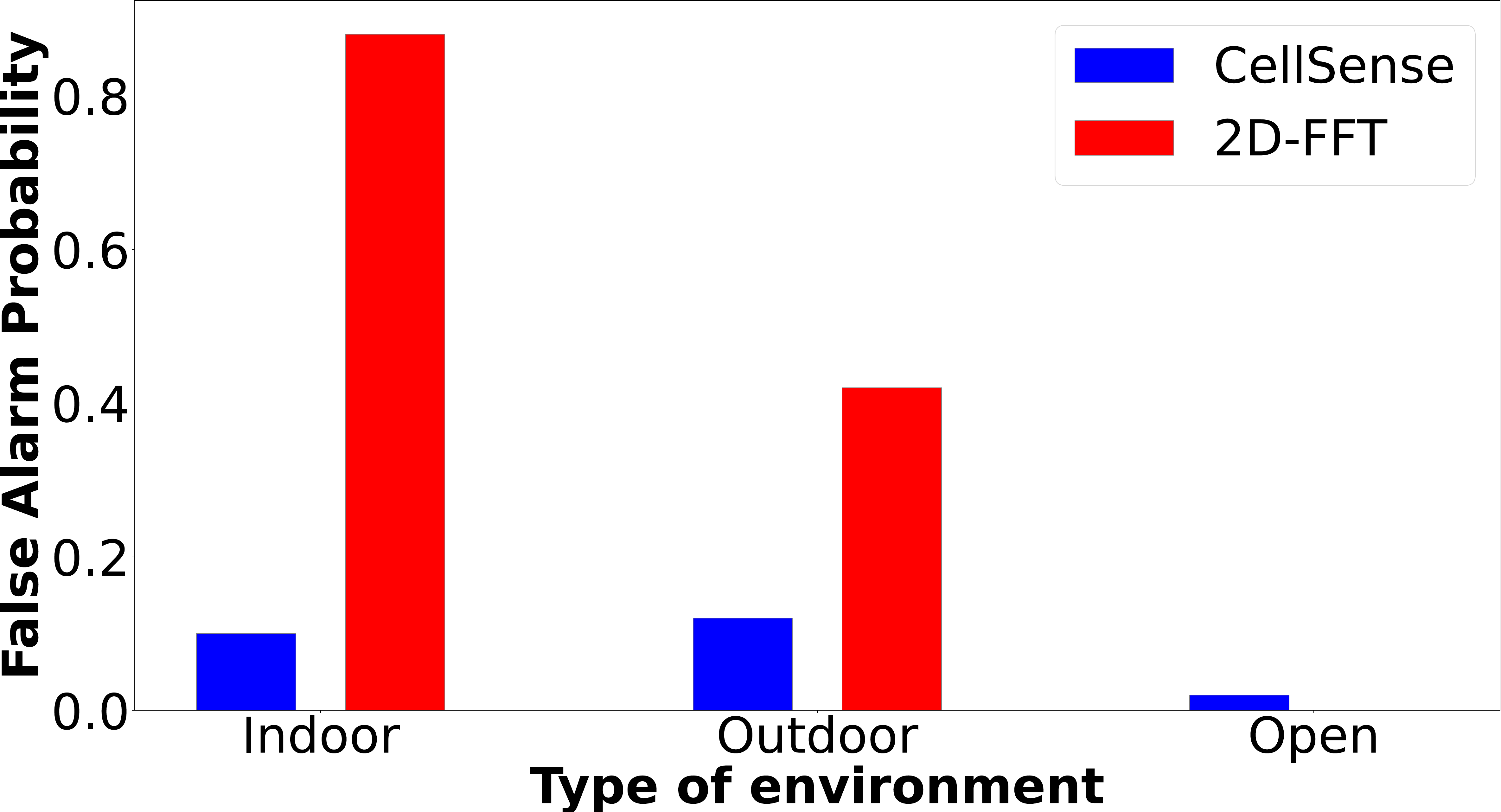}
        \caption{False alarm probability}
        \label{fig:sub_image2}
    \end{subfigure}
    \hfill
    \begin{subfigure}[b]{0.3\textwidth}
        \centering
        \includegraphics[width=\linewidth]{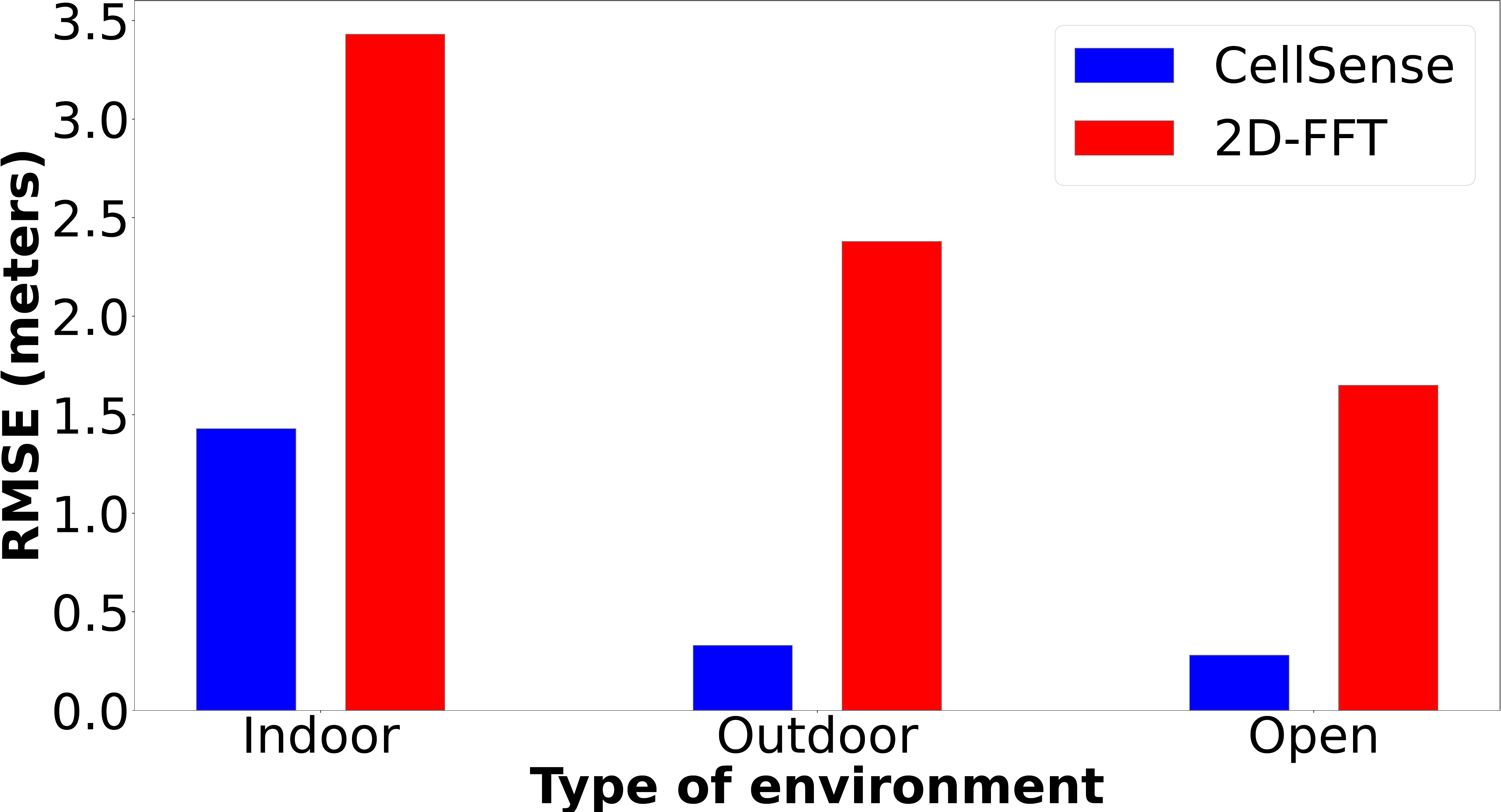}
        \caption{Localization RMSE}
        \label{fig:sub_image3}
    \end{subfigure}
    \vspace{-2mm}
    \caption{Simulation results: Detection and localization performance comparison for \texttt{CellSense} and 2D-FFT}
    %\textcolor{blue}{Increase font size. Add errorbar or replace with boxplot.}}
    \label{fig:bar_graphs}
    \vspace{-5mm}
\end{figure*}

\section{Evaluation}
\texttt{CellSense} is evaluated through a two-fold approach comprising high-fidelity simulations and hardware prototype experiments, and it's performance is benchmarked against a traditional 2D-FFT baseline. The following paragraphs provide details on the baseline implementation, evaluation metrics, and experimental results.

% \textcolor{red}{[I will discuss "Baseline 2D-FFT", "Performance Metrics - Detection Probability, RMSE, and Fall Alarm", and "Three Scenarios - Indoor, Outdoor, and Open" here, as it's applicable to both simulation and hardware study.]}

% \textcolor{red}{[Following this, we can have two subsections - Simulation results and Hardware prototype results.]}

\textbf{Baseline:} To implement the 2D-FFT baseline for object detection, a static reference matrix is first established via cross-correlation of the received signal and the known pilot symbols across the joint AoA-delay search space. This baseline matrix is normalized and stored as a reference. To detect and localize a dynamic object, the stored static 2D-FFT matrix is subtracted from the current normalized 2D-FFT matrix, effectively isolating the dynamic reflections. If the peak magnitude of the resulting differential matrix exceeds a pre-determined threshold, a dynamic object is detected and mapped to the corresponding AoA and transmit delay parameters.

% \textcolor{blue}{Write all details about baseline here. Say that baseline has frame substraction to remove static objects for fairness.}

\textbf{Metrics:} Evaluation relies on three core metrics - \textit{Detection probability} ($P_d$), defined as the ratio of correct detections to the total points where the object is visible to the Base Station (BS); \textit{False alarm probability} ($P_{fa}$), defined as the ratio of false detections to the total algorithm detections; and \textit{Localization Root Mean Square Error (RMSE)}, calculated as the square root of the mean squared error across all estimated target locations.

\subsection{Simulation Experiments} 
For simulation study, we utilize a cross-platform framework: \textit{Blender} is employed to render multiple realistic environments, shown in Fig. \ref{fig:sim_env}, including an indoor warehouse, an outdoor area (Oval at Centennial Campus, NC State University), and an open environment consisting of only the UE, BS, and target on a concrete floor, while the \textit{Sionna ray-tracing library} generates site-specific wireless channel impulse responses. The underlying 5G-compliant OFDM transceiver is implemented in Python, adhering to 3GPP-specified bandwidths and resource configurations. To simulate a dynamic target, we model a mobile object as a $1.8 \ m \times 0.5 \ m \times 0.25 \ m$ cuboid to emulate the average physiological profile of a human. Both the stationary transmitter (UE) and receiver (BS) are positioned at a height of 1 meter. Analysis is performed with the target following multiple constant-velocity trajectories, as shown for the outdoor environment in Fig. \ref{fig:trajectories}. Notably, the target is not visible to the BS for the entire trajectory; detection and localization is only performed for segments where a reflected signal from the object reaches the BS. \par
\textbf{Results:} Detailed system parameters are summarized in Table \ref{table:Simulation Parameters}; these values are used for all simulation analysis. To ensure statistical significance, we conduct 10 Monte Carlo iterations for 3 different trajectories in each scenario. The simulation results are presented in Fig. \ref{fig:bar_graphs}, where ``Indoor'' refers to the indoor warehouse environment, ``Outdoor'' denotes the Oval area at the NCSU Centennial Campus, and ``Open" represents a baseline environment consisting solely of the UE, BS, and target on a concrete floor. These results demonstrate a clear performance advantage for \texttt{CellSense} over the 2D-FFT baseline across all environments. In the high-clutter indoor warehouse, \texttt{CellSense} achieves a detection probability ($P_d$) of 0.74 with a low false alarm rate ($P_{fa}=0.10$), whereas the baseline struggles at $P_d = 0.22$ and a prohibitive $P_{fa} = 0.88$. Localization accuracy follows a similar trend: \texttt{CellSense} maintains sub-meter RMSE in outdoor (0.33 m) and open (0.28 m) settings, while 2D-FFT yields significantly higher errors of 2.38 m and 1.65 m, respectively. Even in complex indoor environments, \texttt{CellSense} delivers an RMSE of 1.43 m, outperforming the baseline by 2 meters. This confirms that \texttt{CellSense}'s high-resolution parameter estimation and clutter suppression provide robust, precise tracking even under unfavorable signal-to-clutter ratios.

\begin{table}[htbp]
\vspace{-0.1in}
\caption{Simulation parameters}
\vspace{-0.1in}
\centering
{\def\arraystretch{1.5}\begin{tabular}{|p{0.30\linewidth}|p{0.10\linewidth}|p{0.30\linewidth}|p{0.10\linewidth}|}

\hline
\textbf{Definition (Notation)} & \textbf{Values} & \textbf{Definition (Notation)} & \textbf{Values} \\
\hline
Carrier frequency ($f_c$) & $3.6192$ $GHz$ & Bandwidth ($B$) & $20$ $MHz$ \\
\hline
Subcarrier spacing ($f_{scs}$) & $30$ $KHz$ & Number of antennas at the BS ($N_{ant}$) & 4\\
\hline
SRS transmission comb size ($K_{TC}$) & 2 & Sampling rate ($f_s$) & $30.72$ $Msps$ \\
\hline
IFFT size ($N_{fft}$) & $1024$ & Number of PRBs ($N_{prb}$) & $51$ \\
\hline
Size of cyclic prefix ($N_{cp}$) & $72$ & Transmit power at UE ($P_t$) & $20$ $dBmW$ \\
\hline
\end{tabular}}
\label{table:Simulation Parameters}
\vspace{-0.2in}
\end{table}

\begin{figure}[t!]
    \centering
    \begin{subfigure}[b]{0.49\columnwidth} 
        \centering
        \includegraphics[width=\linewidth]{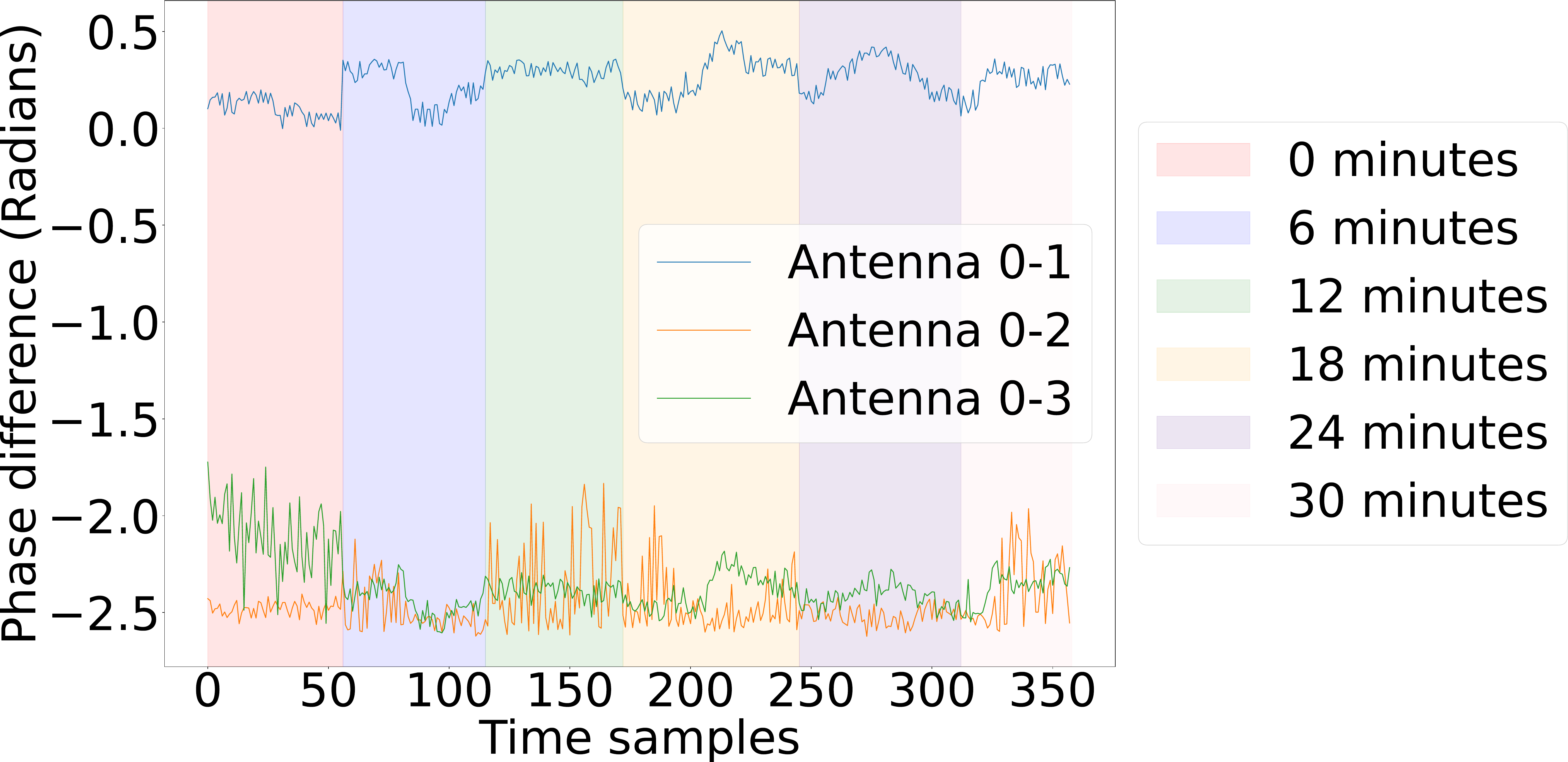}
        \caption{}
        \label{fig:ipo_calibration}
    \end{subfigure}
    \begin{subfigure}[b]{0.49\columnwidth}
        \centering
        \includegraphics[width=\linewidth]{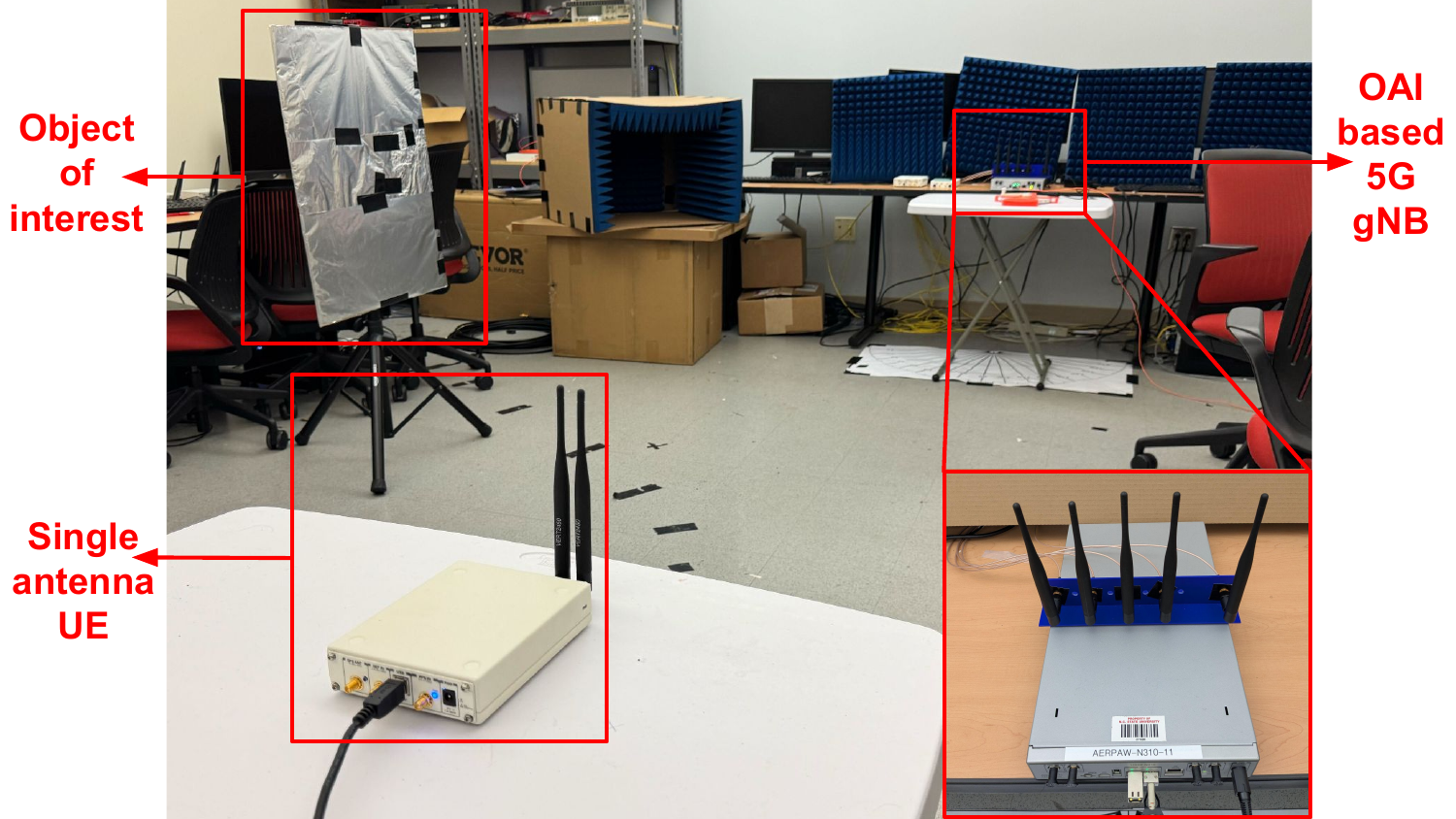}
        \caption{}
        \label{fig:testbed}
    \end{subfigure}
    \vspace{-6mm}
    \caption{Figure (a) indicates variation of phase offset between the RF chains across time. Figure (b) indicates the setup of \texttt{CellSense} hardware prototype.}
    \label{fig:sensing_metrics}
    \vspace{-7mm}
\end{figure}

\subsection{Hardware prototype and results}
\textbf{Hardware setup:} To validate the real-world efficacy of \texttt{CellSense}, we built a hardware testbed (See Fig. \ref{fig:testbed}) utilizing Software Defined Radios (SDRs) in an indoor laboratory environment. The Base Station (BS) is equipped with an Ettus USRP N310 configured with a four-element Uniform Linear Array (ULA) to facilitate high-resolution Angle of Arrival (AoA) estimation, while a single-antenna USRP B210 is employed as the UE. The 5G cellular protocol stack is implemented using the OpenAirInterface (OAI) codebase. The system parameters utilized for hardware evaluation are identical to those specified in Table \ref{table:Simulation Parameters}.\par

\textbf{Calibration:} Multi-antenna SDR systems suffer from an inherent Initial Phase Offset (IPO) between RF chains, requiring baseline calibration for spatial accuracy. As shown in Fig. \ref{fig:ipo_calibration}, our empirical analysis confirms that post-initialization, the IPO remains stable with negligible drift for at least 30 minutes. To validate this across sessions, the UE was power-cycled (disconnected and reconnected) every 6 minutes while the BS ran continuously. Each colored region in the plot corresponds to an individual connection session, representing initial 10 seconds of continuous time samples used to compute the phase difference between the RF chains. For instance, the data points included in the 0-minute region illustrate the localized IPO variation during the initial 10 seconds of the experiment's first session. This stability enables a single initial calibration to maintain accuracy throughout the experiment.

% \begin{figure}[t!]
% \centering
% \includegraphics[width=1\linewidth]{Figures/spatial_results_rmse.pdf}
% \vspace{-0.05in}
% \caption{Visualization of the sensing accuracy and detection probability for CellSense and 2D-FFT for 10 different points spread across the lab environment. \textcolor{blue}{Upload all figures as pdf. Format as subfigure abcd etc. Font size a bit higher. Circle can be bigger in size. }}
% \label{fig:spatial_results}
% \vspace{-0.1in}
% \end{figure}

\begin{figure}[t!]
    \centering
    \begin{subfigure}[b]{1\columnwidth} 
        \centering
        \includegraphics[width=\linewidth]{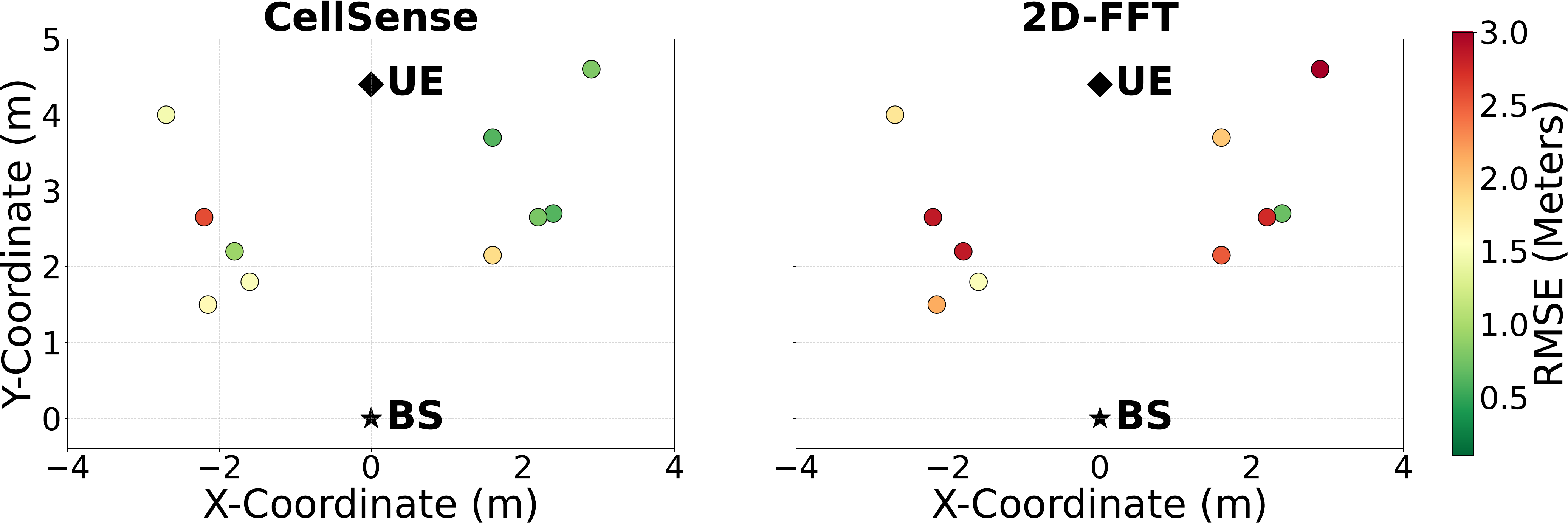}
        \caption{RMSE}
    \end{subfigure}
    \begin{subfigure}[b]{1\columnwidth}
        \centering
        \includegraphics[width=\linewidth]{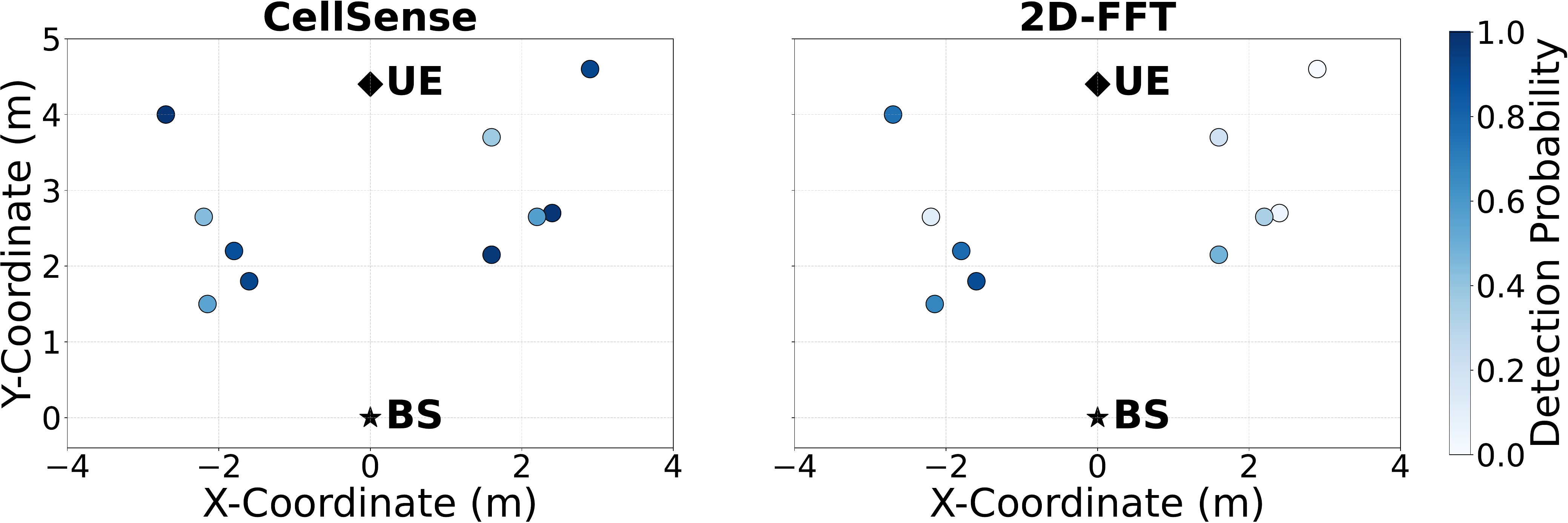}
        \caption{Detection probability}
    \end{subfigure}
    \vspace{-6mm}
    \caption{Spatial sensing accuracy and detection probability across 10 experimental locations. Target ground truth locations are color-coded by (a) RMSE (m) and (b) detection probability (\%) for both \texttt{CellSense} and the 2D-FFT baseline.}
    \label{fig:spatial_results}
    \vspace{-8mm}
\end{figure}

\textbf{Results:} Following the phase calibration, we evaluate \texttt{CellSense}'s spatial performance and communication–sensing trade-off within an indoor laboratory environment (Fig. \ref{fig:testbed}). The static BS and UE are positioned at fixed coordinates to establish the baseline communication link, while the target moves across 10 distinct, pre-mapped experimental locations to evaluate sensing accuracy. This environment presents severe multipath challenges due to the dense presence of strong static specular reflectors, including concrete laboratory walls, metallic equipment racks, and large LED monitor screens. Fig. \ref{fig:spatial_results} maps the spatial Root Mean Square Error (RMSE) and detection probability across these 10 experimental locations. The empirical results demonstrate that \texttt{CellSense} consistently outperforms the 2D-FFT baseline, reducing the average spatial RMSE from 2.2 m to 1.28 m while boosting the mean detection probability from 43\% to 76\%. Notably, at challenging coordinates such as $(2.9, 4.6)$ m and $(2.4, 2.7)$ m, which are located in close proximity to the large monitor screens and laboratory walls, \texttt{CellSense} mitigates the severe clutter and maintains a detection probability exceeding 91\% alongside sub-meter accuracy. Conversely, due to the overwhelming static reflections at these coordinates, the detection probability for the baseline 2D-FFT drops below 10\%. It shows the robustness of \texttt{CellSense} in high-clutter deployment scenarios. We also quantify the communication-sensing tradeoff by varying the SRS periodicity (Table \ref{table:sense_comm_tradeoff}). Tuning the SRS period from a sparse 160 slots (80 ms) to a dense 20 slots (10 ms) improves RMSE from 0.74 m to 0.66 m, incurring a minor throughput penalty of ~25 Kbps. This confirms \texttt{CellSense} achieves reliable, low-latency tracking without sacrificing communication QoS.

\begin{table}[htbp]
\vspace{-0.05in}
\caption{Sensing RMSE vs throughput}
\centering
\vspace{-0.05in}
{\def\arraystretch{1.5}\begin{tabular}{|p{0.25\linewidth}|p{0.25\linewidth}|p{0.25\linewidth}|}

\hline
\textbf{SRS period (slots/ms)} & \textbf{Sensing RMSE (meters)} & \textbf{Throughput (Mbps)} \\
\hline
160/80 & 0.74 & 1.9804 \\
\hline
80/40 & 0.74 & 1.9793 \\
\hline
40/20 & 0.74 & 1.963 \\
\hline
20/10 & 0.66 & 1.953 \\
\hline
\end{tabular}}
\label{table:sense_comm_tradeoff}
\vspace{-0.2in}
\end{table}

\section{Conclusion}

This paper introduced \texttt{CellSense}, a sub-6 GHz ISAC architecture integrated into a functional 5G protocol stack. To overcome high environmental clutter, we proposed a high-resolution SAGE-based multipath estimator paired with a bipartite-matching and sliding-window data association pipeline. Extensive site-specific simulations across indoor warehouse and outdoor campus settings validated high-fidelity tracking. Furthermore, a real-world hardware prototype demonstrated an average indoor localization accuracy of 1.28 m and a 76\% mean detection probability, drastically outperforming the 2D-FFT baseline. Future work will extend this framework to multi-cell cooperative bistatic sensing for expanded coverage and multi-target tracking.

%\vspace{-0.1in}
\bibliographystyle{IEEEtran}
\bibliography{references.bib}

\end{document}